\documentclass[superscriptaddress,showkeys,showpacs,floatfix,aps,prb,twocolumn]{revtex4}

\usepackage{graphicx}
\usepackage{bm}
\usepackage{amsmath}
\usepackage{dcolumn}
\usepackage{multirow}

\newcommand{\ksoc}{$K_{so}$}

\newcommand{\morb}{$M_{l}$}
\newcommand{\kmorb}{$K_{l}$}

\newcommand{\uu}{$\uparrow\uparrow$}
\newcommand{\ud}{$\uparrow\downarrow$}
\newcommand{\du}{$\downarrow\uparrow$}
\newcommand{\dd}{$\downarrow\downarrow$}
\newcolumntype{/}{D{/}{/}{2,2}}  
\newcolumntype{.}{D{.}{.}{0}}  

\input{tcilatex}

\begin{document}

\title[spin-orbit coupling, magnetic anisotropy]{Constituents of magnetic anisotropy and a screening of spin-orbit
  coupling in solids\footnote{Materials of this paper have been presented at the 58th Annual Magnetism and Magnetic Materials
  (MMM) Conference in Denver, Colorado in November 2013.}}
\author{Vladimir Antropov$^{\dagger}$}
\affiliation{Ames Laboratory USDOE, Ames, IA 50011}
\author{Liqin Ke\footnote{Corresponding authors.  E-mail addresses: antropov@ameslab.gov (V. Antropov), liqinke@ameslab.gov (L. Ke).}}
\affiliation{Ames Laboratory USDOE, Ames, IA 50011}
\author{Daniel  \AA{}berg}
\affiliation{Lawrence Livermore National Laboratory, USDOE, Livermore, CA, 94550}
\keywords{D.Magnetic anisotropy; D.Spin-orbit coupling; D.Perturbation; D.Screening}
\pacs{PACS number}

\begin{abstract}
Using quantum mechanical perturbation theory (PT) we analyze how the energy of perturbation of different orders is
renormalized in solids. We test the validity of PT analysis by considering a specific case of spin-orbit coupling
as a perturbation. We further compare the relativistic energy and the magnetic anisotropy from the PT approach with
direct density functional calculations in FePt, CoPt, FePd, MnAl, MnGa, FeNi, and tetragonally strained FeCo. In
addition using decomposition of anisotropy into contributions from individual sites and different spin components
we explain the microscopic origin of high anisotropy in FePt and CoPt magnets.
\end{abstract}
\eid{identifier}
\date{\today }
\maketitle

The magnetocrystalline anisotropy is a central magnetic property for both fundamental and practical
reasons.\cite{{STOHR,kota,kosugi}} It can depend sensitively on many quantities such as dopants or small changes in
lattice constant.\cite{fe8n} While control of this sensitive quantity can be crucial in many applications, e.g.
permanent magnetism\cite{kirill}, magnetooptics\cite{antonov} and magnetoresistive random-access memory
devices\cite{nature}, it is often unclear what mechanisms are responsible for these anisotropy variations, even
from a fundamental point of view. It was understood long ago\cite{ABRA,yosida} that the magnetic anisotropy energy
(MAE) $K$ in bulk materials is a result of simultaneous action of spin-orbit coupling (SOC) and crystal field
(CF). While in general this statement is still valid, existing microscopic methods do not accurately describe $K$
in the majority of materials. One can calculate MAE using \textit{ab-initio} electronic structure methods based on
density functional theory, however quantitative agreement is often rather poor. In any case such methods are
usually not well equipped to resolve it into components that yield an intuitive understanding, to enable its
manipulation and control. Sometimes $K $ is analyzed in terms of SOC matrix elements of ${\xi }\mathbf{l{\cdot
  }s}$, where ${\xi }$ is the SOC constant.  However, this perturbation also induces changes in other terms
contributing the total energy, which can affect the MAE as well. Below we show how the actual atomic SOC is
'screened' in crystals and study spin decomposition of SOC and MAE in real world magnets.

Let us write the total Hamiltonian of magnetic electronic system as%
\begin{equation}
H=H_{0}+V  \label{a1}
\end{equation}%
where $H_{0}$ is the non-relativistic Hamiltonian (sum of kinetic and potential energies of electrons) and $V=\xi
\mathbf{l{\cdot }s}$ is the SOC Hamiltonian. We assume that $\xi $ is small relative to CF and spin splittings.
The change in the total energy of the system when SOC is added (below we call it relativistic part of the total
energy) can be written as
\begin{equation}
 E=\Delta E_{0}+E_{so} \label{a2}
\end{equation}%
where $E_{so}$ is the matrix element of SOC with full perturbed wavefunction and $\Delta E_{0}$ is the induced
energy change of the scalar-relativistic Hamiltonian ( sum of kinetic and potential energies) due to the SOC
perturbation.

Using standard quantum mechanical perturbation theory (PT) each quantity $%
\left\vert \phi \right\rangle =\sum \left\vert n\right\rangle ,E=\sum
E^{\left( n\right) }$ and $E_{so}=\sum V^{\left( n\right) }$ (wave function,
total energy and perturbation $V$) can be expressed as a sum over orders $n$%
: $V^{\left( n\right) }$ is proportional to $\xi ^{n+1}$, while $\left\vert
n\right\rangle $ and $E^{\left( n\right) }$ are of order $\xi ^{n}$.
Here and hereafter we use superscripts in parentheses to denote the order of
perturbation term of the corresponding quantity.
Corresponding expansions can be introduced for the total MAE and MAE due to
SOC term as $K=\sum K^{\left( n\right) }$ and $K_{so}=\sum K_{so}^{\left(
n\right) }$.

If $\left\vert 0\right\rangle $ is an eigenvector of unperturbed system ($%
H_{0}$) then the total perturbation energy can be found as 
\begin{equation}
E=\dsum_{n}E^{\left( n\right) }=\dsum_{n}\left\langle 0\right\vert
V\left\vert n\right\rangle =\left\langle 0\right\vert V\left\vert \phi
\right\rangle   \label{a1a}
\end{equation}%
(see, for instance, Eq.5.1.37 in ref.10 ). It is now straightforward to show
that 
\begin{equation}
E_{so}=\left\langle \phi \right\vert V\left\vert \phi \right\rangle
=\dsum\limits_{n}nE^{\left( n\right) }  \label{a1b}
\end{equation}%
so the sum of kinetic and potential energies change can be presented as 
\begin{equation}
\Delta E_{0}=\dsum\limits_{n}\left( 1-n\right) E^{\left( n\right) }=\left(
\left\langle 0\right\vert -\left\langle \phi \right\vert \right) V\left\vert
\phi \right\rangle   \label{a1c}
\end{equation}

The last expression can be directly evaluated to estimate a reaction of the system to the original perturbation
$V$. In our case this reaction corresponds to joint action of kinetic and potential energy terms ($H_0$ in Eq.\ref
{a1}). Eq.\ref{a1b} is particularly convenient for the analysis due to opportunity to obtain site and spin
decompositions.

Let us consider again a specific case of SOC perturbation $V=\xi \mathbf{l{%
\cdot }s}$ in the second order of PT. In this case we have $E^{\left(
2\right) }=E_{so}^{\left( 1\right) }/2=V^{\left( 1\right) }/2,$ where $%
V^{\left( 1\right) }$ is obtained using wave function of the first order $%
\left\vert \phi \right\rangle \approx $ $\left( \left\vert 0\right\rangle
+\left\vert 1\right\rangle \right) $. Correspondingly, for the second order
MAE 
\begin{equation}
K^{(2)}=K_{so}^{\left( 1\right) }/2  \label{a4}
\end{equation}%
The second order correction to the total MAE due to SOC is a half of the first
order MAE due to SOC only. It is a simple consequence of our perturbation
treatment. One can immediately write down the MAE in cubic systems where the
leading term scales as $\xi ^{4}$, as $K^{(4)}=K_{so}^{\left( 3\right) }/4$.
Thus kinetic and potential terms effectively 'screen' 75\% of the original
SOC MAE in cubic materials. Evidently higher order contributions to total
MAE decrease as $1/n$ relative to SOC anisotropy. Thus the highest anisotropy
can be naturally expected only for a small $n$.

The specific form of the second order correction due to SOC has been studied
many times in different parts of solid state physics\cite{STOHR,ABRA,yosida,LAAN}
and can be obtained if we rewrite $V^{\left( 1\right) }$ as 
\begin{eqnarray}
V^{\left( 1\right) } &=&2\left\langle 0\right\vert \xi \mathbf{l{\cdot }s}%
\left\vert 1\right\rangle =2\xi s_{i}^{\left( 0\right) }l_{i}^{\left(
1\right) }=  \label{a44} \\
&&2\xi s_{i}^{\left( 0\right) }\sum_{{exc}} \frac{\left\langle 0\right\vert
l_{i}\left\vert 1\right\rangle \left\langle 1\right\vert \xi
s_{j}l_{j}\left\vert 0\right\rangle }{\varepsilon'-\varepsilon _{0}} 
\notag \\
&=&2\xi ^{2}s_{i}^{\left( 0\right) }\sum_{exc} \frac{\left\langle 0\right\vert
l_{i}\left\vert 1\right\rangle \left\langle 1\right\vert l_{j}\left\vert
0\right\rangle }{\varepsilon'-\varepsilon _{0}}s_{j}^{\left( 0\right) } 
\notag \\
&=&2\xi s_{i }^{\left( 0\right) }\Lambda _{i j }s_{j }^{\left(
0\right) }  \notag
\end{eqnarray}%
where we indicated the specific orders for spin and orbital moments
entering $V^{\left( 1\right) }$, $\varepsilon_0$ and $\varepsilon'$
are ground state and excited state energy respectively for the
unperturbed system, and the sum is over all excited states. The
leading relativistic correction for the spin moment $s$ appears only
in the second order in $\xi $ and does not contribute to $V^{\left(
  1\right) }$. Below we assume that $s$ does not change from its zero
order value. This result (Eq.\ref{a44}) is the familiar expression
\cite{ABRA,yosida,LAAN,STOHR} for the second order spin Hamiltonian
due to SOC, where orbital moment tensor $\Lambda =$ $l^{\left(
  1\right) }/\xi s$ . Correspondingly, in the uniaxial system
(assuming $\Lambda _{\nu \mu }$ is diagonal) we have $K^{(2)}=\xi
s\left( l_{z}^{\left( 1\right) }-l_{x}^{\left( 1\right) }\right) =\xi
^{2}s^{2}(\Lambda _{\perp }-\Lambda _{\parallel }).$

One can regard the total relativistic energy as the energy change due to the \textquotedblleft
atomic\textquotedblright\ SOC (i.e. matrix elements of ${ \xi }\mathbf{l{\cdot }s}$), `screened' or reduced by
adjustments in other contributions to the total energy. The same evidently holds true for the total relativistic
energy change relative to SOC energy alone even in the nonmagnetic case. One can rewrite Eq.\ref{a2} as
\begin{equation}
E=\Delta E_{0}+\left\langle \xi \mathbf{l{\cdot }s}\right\rangle
=\langle \widetilde{\xi }\mathbf{l{\cdot }s}\rangle  \label{a5}
\end{equation}%
where $\widetilde{\xi }$ is a screened or effective crystal SOC constant as opposed to the atomic or
nonrenormalized $\xi $. We call $\widetilde{\xi }/\xi $ ratio \textit{spin-orbit reduction} factor. One can compare
this parameter with the enhancement of SOC discussed in Ref.~\onlinecite{yafet}.

According to above results (Eq.\ref{a1b}) $\widetilde{\xi }=\xi /2$ (second
order correction) and $\widetilde{\xi }=\xi /4$ (fourth order correction).
Thus the effective screening is minimal for systems with large SOC and
non-cubic symmetries. Evidently this conclusion supports traditionally large
anisotropies observed in magnetic uniaxial systems.

Thus $H_{0}$ term in Eq.\ref{a1}, the sum of kinetic and potential energies,
reduces the effect of SOC and makes overall strength twice smaller in second
order, so $K_{kin}{+}K_{pot}{=}-K_{so}/2$. Overall the action of these terms
is destructive for materials with observed uniaxial anisotropy as total $K$
is opposite in sign to the anisotropy induced by kinetic and potential terms
together $K=-\left( K_{kin}{+}K_{pot}\right) $. Also comparing Eq.\ref{a1b}
and Eq.\ref{a1c} one can see that for arbitrary $n$ ratio $E_{so}^{\left(
n\right) }/E_{0}^{\left( n\right) }=n/\left( 1-n\right) $, thus for large $n$
this ratio tends to be equal to $-1$ meaning that SOC effects are nearly
completely screened in this limit.




\begin{table}[ht]
\caption{ $c/a$ ratio( with respect to the primitive cell), calculated {$K$} and {$K_{so}$}/{$K$} ratio in uniaxial
  magnetic systems. For all systems experimental structures have been used, while for FeCo , we used hypothetical
  tetragonally strained structure. }
\label{tbl1}
\begin{tabular}{cccc}
\hline
Compounds & $c/a$   & $K$($\mu eV$/f.u.)& {$K_{so}$}/{$K$} \\ \hline
FePt      &  1.362  &   2661    &     1.84         \\
CoPt      &  1.379  &    837    &     1.67         \\
FeNi      &  1.414  &     87    &     1.98         \\
FePd      &  1.370  &    174    &     2.14         \\
MnAl      &  1.294  &    287    &     1.98         \\
MnGa      &  1.280  &    437    &     1.99         \\
FeCo      &  1.1    &    216    &     2.21         \\
\hline
\end{tabular}
\end{table}

Let us now consider electronic structure calculations for realistic systems. Using the Vienna ab initio simulation
package\cite{vasp} method we obtained the relativistic energy $E=(E_{r}-E_{nr})$ and SOC energy $E_{so}$ in
non-magnetic and magnetic systems, where $E_{nr}$ and $E_{r}$ are total energies obtained in scalar relativistic
and calculations where SOC has been added (relativistic). The SOC is included \cite{anderson} using the
second-variation procedure. The generalized gradient approximation of Perdew, Burke, and Ernzerhof was
used for the correlation and exchange potentials.  The nuclei and core electrons were described by projector
augmented wave potentials and the wave functions of valence electrons were expanded in a plane-wave basis set
with a energy cutoff between 348 eV and 368 eV for all compounds we investigated in this work. The $k$-point
integration was performed using a tetrahedron method with Bl\"ochl corrections with 13800 $k$-points in the first
Brillouin zone corresponding to the primitive unit cell of $L1_0$ structure.

We compared the spin-orbit reduction factor $\alpha {=}E_{so}/E$ for Al and non-magnetic Fe. The resulting $\alpha
$ appears to be very close to 2 with small deviations of about 1-3$\%$. For magnetic systems, we also found that $
\alpha {\approx }2$ for different magnetization directions.

MAE in $L1_0$ compounds and tetragonal FeCo had been well
studied\cite{kota,kosugi,ravindran,solovyev,sakuma,burkert}.  The calculated MAE values are in reasonable agreement
with previous calculations\cite{kota,kosugi}. For CoPt, the discrepancy between current calculation and previous
ones is rather large. This is due to the exchange correlation potential used, our LDA calculation gives a MAE
about 1.3$meV$/f.u., which is in better agreement with previous calculations.

Here we investigated $K_{so}/K$ in those systems and the results are
presented in Table \ref{tbl1}. The anisotropic part of $E_{so}$ appears to be much smaller than the isotropic part,
and deviations of $K_{so}$/$K$ from 2 are already significant. For instance $ K_{so}$/ $K$ in CoPt is 1.67-1.8
depending on the exchange-correlation potential used.  Compared with $E_{so}$, $K_{so}$ is a much smaller quantity.
The deviations of $K_{so}$/$K$ from the factor two in table 1 are related to a deviation from a second order PT.
That includes both self-consistency effects and a contribution from higher order terms of PT.



\begin{table*}[ht]
\caption{Spin decomposition of atomic spin-orbit coupling energy $E_{so}$ (meV) and orbital magnetic moment
  {$M_{l}$} ($10^{-3}\mu_{B}$). {$\uparrow\uparrow$} indicates the majority spin channel, and {$
  \downarrow\downarrow$} indicates the minority spin channel. Anisotropies of $ E_{so}$ and {$M_{l}$} are defined
as {$K_{so}$}=$( E_{so}^x-E_{so}^z )$ and { $K_{l}$}=$(M_{l}^z-M_{l}^x)$ respectively.}
\label{tbl2}%
\begin{tabular}{c|c|ccccc|ccccc}
\hline
\hline
\multicolumn{2}{c|}{\multirow{2}{*}{FePt}}        & \multicolumn{5}{c|}{Pt}                                           & \multicolumn{5}{c}{Fe}                          \\
\multicolumn{2}{c|}{}        &      {\uu}   &    {\ud}   &    {\du}   &    {\dd}   &    {Total} &    {\uu}    &   {\ud}   &    {\du}   &    {\dd}  &    {Total}  \\
\hline
               &    $z$      &    -133.80   &  -281.85   &  -281.85   &  -145.66   &  -843.2    &    -1.08    &   -2.47   &    -2.47   &    -2.62  &    -8.6   \\
$E_{so}$      &     $x$      &    -125.88   &  -282.87   &  -282.87   &  -145.22   &  -836.9    &    -1.53    &   -2.86   &    -2.86   &    -2.83  &   -10.1   \\
               &    {\ksoc}  &       7.92   &    -1.02   &    -1.02   &     0.44   &     6.3    &    -0.45    &   -0.38   &    -0.38   &    -0.21  &    -1.5   \\
\hline                                                                                                                                         
               &    $z$      &      -113    &            &            &     166    &     53     &    -13      &           &            &       77  &      64   \\
{\morb}      &    $x$      &       -98    &            &            &     165    &     67     &    -23      &           &            &       85  &      62   \\
               &    {\kmorb} &       -15    &            &            &       1    &    -14     &     10      &           &            &       -8  &       2   \\
\hline
\hline
\multicolumn{2}{c|}{\multirow{2}{*}{CoPt}}        & \multicolumn{5}{c|}{Pt}                                           & \multicolumn{5}{c}{Co}                          \\
\multicolumn{2}{c|}{}        &      {\uu}   &    {\ud}   &    {\du}   &    {\dd}   &    {Total} &    {\uu}    &   {\ud}   &    {\du}   &    {\dd}  &    {Total}  \\
\hline
               &    $z$      &    -134.46   &  -287.69   &  -287.69   &  -151.16   &  -861.0    &    -1.40   &    -2.97   &    -2.97   &    -4.97  &   -12.3   \\
$E_{so}$       &    $x$      &    -126.18   &  -290.64   &  -290.64   &  -151.30   &  -858.8    &    -1.94   &    -3.57   &    -3.57   &    -4.08  &   -13.2   \\
               &    {\ksoc}  &       8.28   &    -2.95   &    -2.95   &    -0.14   &     2.2    &    -0.55   &    -0.60   &    -0.60   &     0.90  &    -0.90  \\
\hline                                                                               
               &    $z$      &      -112    &            &            &     185    &     72     &    -16     &            &            &      116  &    100    \\
{\morb}      &    $x$      &       -96    &            &            &     188    &     91     &    -26     &            &            &       92  &     66    \\
               &    {\kmorb} &       -16    &            &            &      -3    &    -19     &     10     &            &            &       24  &     34    \\
\hline
\hline

\end{tabular}%
\end{table*}

It had often been discussed\cite{STOHR,yosida,streever} how a minimum of energy is related to a maximum of
orbital magnetic moment {$M_{l}$}. However, in realistic systems such a relation is often not fulfilled.  For
example, FePt and CoPt have smaller orbital magnetic moments along the easy axis\cite{sakuma}. In order to
understand these phenomena, we resolve SOC\ energies and orbital moments into atomic and spin contributions for FePt
and CoPt (Table \ref{tbl2}). {$ M_{l}$} contains two spin longitudinal contributions $M_{l}=-L_{z}^{\uparrow
  \uparrow }+L_{z}^{\downarrow \downarrow }$, while the energy {$ E_{so}^{\left( 1\right) }$}${\sim }-\xi /2\left(
L_{z}^{\uparrow \uparrow }+L_{z}^{\downarrow \downarrow }+L_{-}^{\uparrow \downarrow }+L_{+}^{\downarrow \uparrow
}\right) $ also contains spin transversal terms. As shown in Table \ref{tbl2}, Pt atoms produce the dominant
contribution to {$K_{so}$}, while Fe (Co) atoms have much smaller and negative values. For each spin component
on Pt atom a direct proportionality between the orbital moment and the SOC energy is
confirmed, thus a larger component of orbital moment corresponds to the minimum of corresponding energy
component. However, this is no longer true for the total moment due to the fact that term with $L_{z}^{\downarrow
  \downarrow }$ enters definition of total moment and energy with different sign (see above). In addition, as shown
in Table \ref{tbl2}, while {$\downarrow \downarrow $} components of {$ M_{l}$} and {$E_{so}$ are} large but rather
isotropic, their {$\uparrow \uparrow $} components are smaller but much more anisotropic, providing a dominant
contribution to the total MAE.



\begin{figure}[ht]
\begin{tabular}{c}
\includegraphics[width=.45\textwidth,clip]{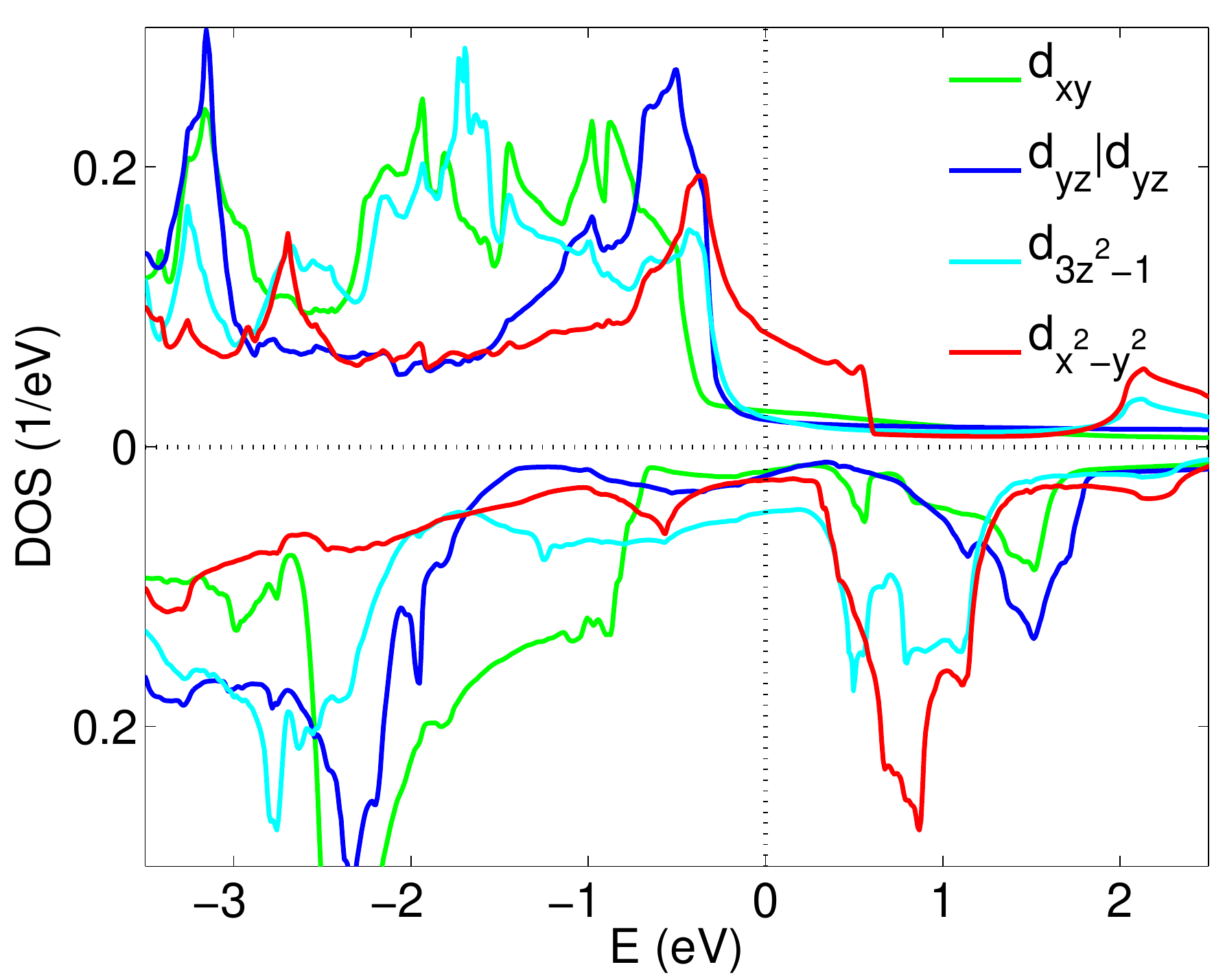} \\
\includegraphics[width=.45\textwidth,clip]{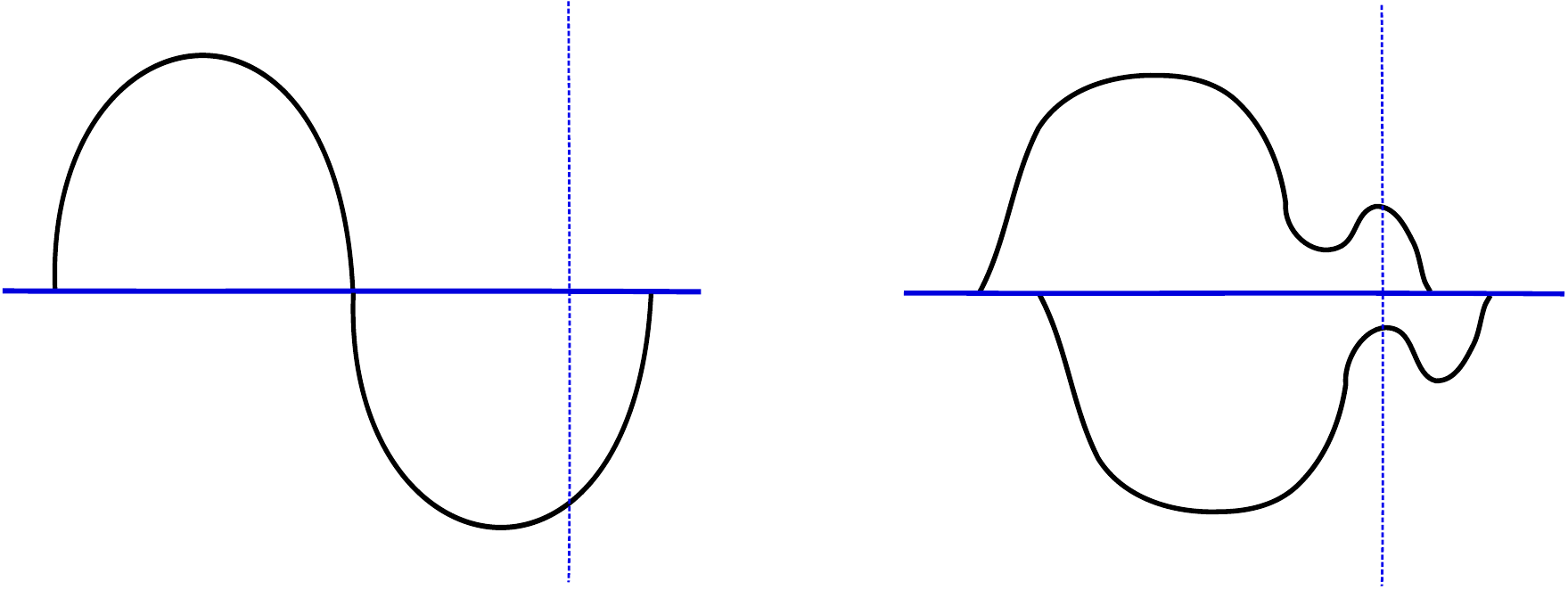} \\
\end{tabular}
\caption{(Color online) Partial density of state (top) projected on the $5d$ states of Pt atom in FePt and
  schematic (bottom) representation of DOS of an ideal nearly-filled $d$-band and FePt. The vertical dotted line
  corresponds to the Fermi energy, $E_F$. }
\label{fig3}
\end{figure}

A different behavior of anisotropies of these two spin channels on Pt sites we attribute to the peculiar features
in the density of states (DOS). Fig.\ref{fig3} (top) shows the partial DOS projected on the $5d$ states of Pt atom
in FePt and CoPt. For the majority spin channel, there are large DOS of all $5d$ states right below the Fermi level
and also a large density of state of $d_{x^{2}-y^{2}}$ right at the Fermi level, which is unexpected for elements
with nearly filled $d$-band. The minority spin channels have smaller DOS right below the Fermi level, especially
for the $d_{x^{2}-y^{2}}$ and degenerated $d_{yz}|d_{xz}$ states. Fig.\ref{fig3} (bottom) shows the schematic
representation of difference between FePt DOS and an ideal nearly-filled $d$-band DOS with a large spin-splitting.
Hence we would expect the minority spin channel has smaller contribution to MAE than the majority one.


In this paper using general perturbation theory we discuss how the spin-orbit interaction is renormalized in
solids. We show that kinetic and potential energy terms nearly completely 'screen' spin-orbit coupling at higher
orders of perturbation theory. By decomposing the MAE and atomic orbital moments into the sum over different spin
matrix elements, we explained why FePt and CoPt have smaller orbital magnetic moments along the easy axis and a
microscopic source of large anisotropy in these materials. Such analysis of the SOC energy makes it easier to study
atomic decomposition of MAE and other anisotropic effects. 

This research is supported in part by the Critical Materials Institute, an Energy Innovation Hub funded by the
U.S. Department of Energy, Office of Energy Efficiency and Renewable Energy, Advanced Manufacturing Office and by
its Vehicle Technologies Program, through the Ames Laboratory. LLNL is operated under Contract DE-AC52-07NA27344
and Ames Laboratory is operated by Iowa State University under contract DE-AC02-07CH11358.

\end{document}